\documentclass[a4paper,11pt]{article}
\pdfoutput=1 

\usepackage{jheppub} 
\usepackage{lineno}
\usepackage{orcidlink}
\usepackage{hhline}
\usepackage{slashed}
\usepackage{graphicx}
\usepackage{graphics}
\usepackage{dcolumn}
\usepackage{bm}
\usepackage{amssymb}
\usepackage{enumerate}
\usepackage{lineno}
\usepackage{multirow} 
\usepackage{color}
\usepackage{hyperref}

\newcommand*{\Zpr}{{Z^\prime}}
\newcommand*{\Apr}{{A^\prime}}
\arxivnumber{2404.06982} 
\title{First constraints on the $L_\mu-L_\tau$  explanation of the muon g-2
anomaly from NA64-$e$ at CERN}

\author[a]{Yu.~M.~Andreev\orcidlink{0000-0002-7397-9665}}
\author[b]{A.~Antonov\orcidlink{0000-0003-1238-5158}}
\author[c]{D.~Banerjee\orcidlink{0000-0003-0531-1679}}
\author[d]{B.~Banto Oberhauser\orcidlink{0009-0006-4795-1008}}
\author[c]{J.~Bernhard\orcidlink{0000-0001-9256-971X}}
\author[b,e]{P.~Bisio\orcidlink{/0009-0006-8677-7495}}
\author[b]{A.~Celentano\orcidlink{0000-0002-7104-2983}}
\author[c]{N.~Charitonidis\orcidlink{0000-0001-9506-1022}}
\author[f]{D.~Cooke}
\author[d]{P.~Crivelli\orcidlink{0000-0001-5430-9394}}
\author[d]{E.~Depero\orcidlink{0000-0003-2239-1746}}
\author[a]{A.~V.~Dermenev\orcidlink{0000-0001-5619-376X}}
\author[a]{S.~V.~Donskov\orcidlink{0000-0002-3988-7687}}
\author[a]{R.~R.~Dusaev\orcidlink{0000-0002-6147-8038}}
\author[g]{T.~Enik\orcidlink{0000-0002-2761-9730}}
\author[g]{V.~N.~Frolov}
\author[h]{A.~Gardikiotis\orcidlink{0000-0002-4435-2695}}
\author[a]{S.~N.~Gninenko\orcidlink{0000-0001-6495-7619}}
\author[i]{M.~H\"osgen}
\author[a]{V.~A.~Kachanov\orcidlink{0000-0002-3062-010X}}
\author[g]{Y.~Kambar\orcidlink{0009-0000-9185-2353}}
\author[a]{A.~E.~Karneyeu\orcidlink{0000-0001-9983-1004}}
\author[g]{G.~Kekelidze\orcidlink{0000-0002-5393-9199}}
\author[i]{B.~Ketzer\orcidlink{0000-0002-3493-3891}}
\author[a]{D.~V.~Kirpichnikov\orcidlink{0000-0002-7177-077X}}
\author[a]{M.~M.~Kirsanov\orcidlink{0000-0002-8879-6538}}
\author[a]{V.~N.~Kolosov}
\author[g]{S.~V.~Gertsenberger\orcidlink{0009-0006-1640-9443}}
\author[c]{S.~Girod}
\author[g]{E.~A.~Kasianova}
\author[a,g]{V.~A.~Kramarenko\orcidlink{0000-0002-8625-5586}}
\author[a]{L.~V.~Kravchuk\orcidlink{0000-0001-8631-4200}}
\author[a,g]{N.~V.~Krasnikov\orcidlink{0000-0002-8717-6492}}
\author[l,m]{S.~V.~Kuleshov\orcidlink{0000-0002-3065-326X}}
\author[a,m,n]{V.~E.~Lyubovitskij\orcidlink{0000-0001-7467-572X}}
\author[g]{V.~Lysan\orcidlink{0009-0004-1795-1651}}
\author[b]{A.~Marini\orcidlink{0000-0002-6778-2161}}
\author[b]{L.~Marsicano\orcidlink{0000-0002-8931-7498}}
\author[g]{V.~A.~Matveev\orcidlink{0000-0002-2745-5908}}
\author[m]{R.~Mena~Fredes}
\author[m,n]{R.~Mena~Yanssen}
\author[o]{L.~Molina Bueno\orcidlink{0000-0001-9720-9764}}
\author[d]{M.~Mongillo\orcidlink{0009-0000-7331-4076}}
\author[g]{D.~V.~Peshekhonov\orcidlink{0009-0008-9018-5884}}
\author[a]{V.~A.~Polyakov\orcidlink{0000-0001-5989-0990}}
\author[p]{B.~Radics\orcidlink{0000-0002-8978-1725}}
\author[g]{K.~Salamatin\orcidlink{0000-0001-6287-8685}}
\author[a]{V.~D.~Samoylenko}
\author[d]{H.~Sieber\orcidlink{0000-0003-1476-4258}}
\author[a]{D.~Shchukin\orcidlink{0009-0007-5508-3615}}
\author[m,q]{O.~Soto}
\author[a]{V.~O.~Tikhomirov\orcidlink{0000-0002-9634-0581}}
\author[a]{I.~Tlisova\orcidlink{0000-0003-1552-2015}}
\author[a]{A.~N.~Toropin\orcidlink{0000-0002-2106-4041}}
\author[o]{M.~Tuzi\orcidlink{0009-0000-6276-1401}}
\author[l]{P.~Ulloa\orcidlink{0000-0002-0789-7581}}
\author[a,g]{P.~V.~Volkov\orcidlink{0000-0002-7668-3691}}
\author[a]{V.~Yu.~Volkov\orcidlink{0009-0005-3500-5121}}
\author[a]{I.~V.~Voronchikhin\orcidlink{0000-0003-3037-636X}}
\author[l,m]{J.~Zamora-Sa\'a\orcidlink{0000-0002-5030-7516}}
\author[g]{A.~S.~Zhevlakov\orcidlink{0000-0002-7775-5917}}

\affiliation[a]{Authors affiliated with an institute covered by a cooperation agreement with CERN}
\affiliation[b]{INFN, Sezione di Genova, 16147 Genova, Italia}
\affiliation[c]{CERN, European Organization for Nuclear Research, CH-1211 Geneva, Switzerland}
\affiliation[d]{ETH Z\"urich, Institute for Particle Physics and Astrophysics, CH-8093 Z\"urich, Switzerland}
\affiliation[e]{Universit\`a degli Studi di Genova, 16126 Genova, Italia}
\affiliation[f]{UCL Departement of Physics and Astronomy, University College London, Gower St. London WC1E 6BT, United Kingdom}
\affiliation[g]{Authors affiliated with an international laboratory covered by a cooperation agreement with CERN}
\affiliation[h]{Physics Department, University of Patras, 265 04 Patras, Greece}
\affiliation[i]{Universit\"at Bonn, Helmholtz-Institut f\"ur Strahlen-und Kernphysik, 53115 Bonn, Germany}
\affiliation[l]{Center for Theoretical and Experimental Particle Physics, Facultad de Ciencias Exactas, Universidad Andres Bello, Fernandez Concha 700, Santiago, Chile}
\affiliation[m]{Millennium Institute for Subatomic Physics at High-Energy Frontier (SAPHIR), Fernandez Concha 700, Santiago, Chile}
\affiliation[n]{Universidad T\'ecnica Federico Santa Mar\'ia and CCTVal, 2390123 Valpara\'iso, Chile}
\affiliation[o]{Instituto de Fisica Corpuscular (CSIC/UV), Carrer del Catedratic Jose Beltran Martinez, 2, 46980 Paterna, Valencia, Spain}
\affiliation[p]{Department of Physics and Astronomy, York University, Toronto, ON, Canada}
\affiliation[q]{Departamento de Fisica, Facultad de Ciencias, Universidad de La Serena, Avenida Cisternas 1200, La Serena, Chile}

\emailAdd{luca.marsicano@ge.infn.it}

\abstract{
The inclusion of an additional $U(1)$ gauge $L_\mu-L_\tau$ symmetry would release the tension between the measured and the predicted value of the anomalous muon magnetic moment: this paradigm  assumes the existence of a  new, light $Z^\prime$ vector boson, with dominant coupling to $\mu$ and $\tau$ leptons and interacting with electrons via a loop mechanism. The $L_\mu-L_\tau$ model can also explain the Dark Matter relic abundance, by assuming that the $\Zpr$ boson acts as a ``portal''  to a new Dark Sector of particles in Nature, not charged under known interactions. 
In this work we present the results of the $\Zpr$ search performed by the NA64-$e$ experiment at CERN SPS, that  collected $\sim 9\times10^{11}$ 100 GeV electrons impinging on an active thick target.  Despite the suppressed $\Zpr$ production yield with an electron beam, NA64-$e$ provides the first accelerator-based results excluding the $g-2$ preferred band of the $\Zpr$ parameter space in the 1 keV $ < m_\Zpr \lesssim 2$ MeV range, in complementarity with the limits recently obtained by the  NA64-$\mu$ experiment with a muon beam.

}

\begin{document}

\maketitle

\section{Introduction}
The Standard Model (SM) of fundamental interactions is one the greatest successes of  particle physics, explaining many phenomena at different energy scales. Despite these results, the SM needs to be extended to account for several experimentally observed anomalies or effects, currently not described by the model. A remarkable example is provided by the Dark Matter (DM) particle content puzzle. Nowadays, the existence of DM is confirmed by multiple astrophysical and cosmological observations, but the SM does not include any viable DM particle candidate~\cite{Bertone:2004pz,Arcadi:2017kky,Roszkowski:2017nbc}. This calls for an extension of the SM, with new fields and forces not yet experimentally observed~\cite{Zurek:2024qfm}. Another significant example is provided by the measurement of the anomalous muon magnetic moment $a_\mu \equiv (g_\mu-2)/2$. For this observable, the most updated experimental average, mostly constrained by the latest Fermilab Muon $g-2$ experiment, reads $a_\mu\mathrm{(Exp)}=116\,592\,059(22)\times10^{-11}$~\cite{Muong-2:2023cdq}, to be compared with the latest theoretical prediction from the Muon $g-2$ Theory Initiative, $a_\mu\mathrm{(Teo)}=116\,591\,810(43)\times10^{-11}$~\cite{Aoyama:2020ynm}. The leading order hadronic contribution for $a_{\mu}$ comes from the cross section for the $e^+e^-\rightarrow$~hadrons process, measured by many experiments. While new results from both the experimental side (e.g., the latest CMD-3 result~\cite{CMD-3:2023alj}, still unpublished) and the phenomenological one (e.g., a recent lattice calculation by the BMW collaboration~\cite{Borsanyi:2020mff}) tend to reduce the experiment-to-theory discrepancy, still this difference motivates the investigation of new physics scenarios with a preferred connection to SM second generation leptons.

One of the most attractive and simplest model explaining the $g-2$ discrepancy, the so-called $L_\mu-L_\tau$ scenario, predicts the existence of a new $U(1)$ gauge boson, $Z^\prime$, that couples to second and third generation leptons via a coupling $g_\Zpr$~\cite{He:1990pn,He:1991qd}, and can also be connected to DM phenomenology~\cite{Baek:2022ozm}.
Although cosmological and astrophysical considerations put stringent constraints on the model parameter space, see e.g. Ref.~\cite{Escudero:2019gzq, Kamada:2015era},  so far no laboratory-based experiment explored the preferred parameters region releasing the $g-2$ tension in the sub-MeV $\Zpr$ mass range. Recently, NA64-$\mu$ at CERN  set its first accelerator-based constraints using a muon beam~\cite{NA64:2024klw}, excluding the $L_\mu-L_\tau$ explanation of the  $g-2$ anomaly for $\Zpr$  masses heavier than $\sim20$ MeV. In complementarity with this result, we present in  this work the first results  from the electron-beam missing-energy experiment NA64-e,  wich exclude the $g-2$ band in the 1 keV $ < m_\Zpr \lesssim 2$ MeV mass range.
We note that, in principle, the NA64-$e$ experiment could be sensitive also to a $\Zpr$ lighter than 1 keV, but the production of such a light particle would be affected by non trivial wave function coherence-loss effects~\cite{Demidov_2019}, whose evaluation goes beyond the aim of this work. 
These results are based on a $\sim$3-times larger accumulated statistics if compared to the previous limits set by NA64-$e$~\cite{NA64:2022rme}, resulting in a significant improvement of the explored parameter space. Compared to the previous work, the analysis has been refined by performing a full signal simulation via the DMG4 package~\cite{Bondi:2021nfp,Oberhauser:2024ozf}, including an effective description of the atomic electrons motion in the target. In addition, the limits have been obtained with a rigorous statistical approach,  improving the uncertainty in the determination of the systematic effects.

\section{The $L_\mu-L_\tau$ model} This model introduces the following new lagrangian terms~\cite{Bauer:2018onh}:
\begin{align}
    \mathcal{L} \subset & -\frac{1}{4}\Zpr_{\mu\nu}{\Zpr}^{\mu\nu} +\frac{1}{2}m^2_\Zpr \Zpr_\mu \Zpr^\mu  \\ \nonumber
    & -  g_\Zpr \Zpr_\mu \left (\overline{\mu}\gamma^\mu\mu +\overline{\nu}_\mu\gamma^\mu P_L \nu_\mu  
    -\overline{\tau}\gamma^\mu\tau
    -\overline{\nu}_\tau\gamma^\mu P_L \nu_\tau  
    \right) \;\;,
\end{align}
where $Z^\prime_{\mu\nu}\equiv\partial_\mu \Zpr_\nu-\partial_\nu \Zpr_\mu$ is the $\Zpr$ field strength, $m_\Zpr$ is the $\Zpr$ mass, and $P_L=(1-\gamma_5)/2$.  
In this model, loop-order effects generate an additional positive contribution to $a_\mu(\mathrm{Theo})$, bringing it closer to $a_\mu(\mathrm{Exp})$~\cite{PhysRevD.64.055006}. In particular, for values $g_\Zpr~\simeq~4.5\times10^{-4}$ and $m_\Zpr \ll m_\mu$, the $Z^\prime$ contribution would actually solve the muon $g-2$ discrepancy. Recently, results from the search of a $\Zpr$ boson decaying to muons have been reported by the BaBar~\cite{BaBar:2016sci}, Belle~\cite{Belle:2021feg}, Belle-II~\cite{Belle-II:2024wtd}, and CMS~\cite{CMS:2018yxg} experiments, with null observations so far. The same result was reported by Belle-II and BESIII from the search of an invisibly-decaying $\Zpr$~\cite{Belle-II:2022yaw,BESIII:2023jji}.

The model can also be connected to the DM phenomenology in the context of Dark Sector (DS) scenarios, by postulating the existence of a new ensemble of particles in Nature, not charged under SM interactions. The lightest particle  of the DS (here denoted as $\chi$), if stable, can be the DM candidate. Assuming, for illustration, a complex scalar nature for the $\chi$, the Lagrangian is extended by the $\Zpr-\chi$ interaction term ${\cal L}_{D} = g_D \Zpr_\mu J^\mu_D$, where $J^\mu_D=\left( \chi^* i\partial^\mu \chi - \chi i\partial^\mu \chi^* \right)$ is the DS current and $g_D\equiv\sqrt{4\pi\alpha_D}$ is the  coupling constant. In this picture, considering the mass hierarchy $m_{Z^\prime}>m_\chi$, and far from the resonance condition $m_{Z^\prime}=2m_\chi$, the presently observed DM relic abundance is set by the velocity-averaged cross section for the process $l\overline{l} \rightarrow Z^\prime \rightarrow \chi \chi$, where $l$ is a SM lepton, if DM had a thermal origin in the early Universe. This results to a preferred combination of the model parameters, related as~\cite{Kahn:2018cqs}:
\begin{equation}\label{eq:bound}
    {g^2_D} g^2_\Zpr \left(\frac{m_\chi}{m_\Zpr}\right)^4 \simeq  3 \cdot 10^{-15}\left(\frac{m_\chi}{1\,\mathrm{MeV}} \right)^2\; \; .
\end{equation}

As discussed in Ref.~\cite{NA64:2022rme}, even if the $L_\mu - L_\tau$ model does not explicitly include a tree-level coupling between electrons and the $Z^\prime$, this is straightforwardly introduced via the one-loop level kinetic mixing between the $Z^\prime$ and the photon, resulting to the effective coupling $e \Pi(q^2)$ dependent on the momentum squared $q^2$ carried by the $\Zpr$~\cite{Araki:2017wyg,Gninenko:2018tlp,Zhang:2020fiu}\footnote{The full analytic expression for $\Pi(q^2)$ can be found in the Appendix of Ref.~\cite{NA64:2022rme}.}:
\begin{equation}
\label{eq:pi}
    \Pi(q^2)=\frac{e\,g_\Zpr}{2\pi^2}\int_0^1dx \, x (1-x) \ln\frac{m^2_\tau-x(1-x)q^2}{m^2_\mu-x(1-x)q^2}\;\;.
\end{equation}
In particular, for $q^2 \ll m^2_\mu$, the  loop function $\Pi(q^2)$ approaches to $\Pi(q^2)\simeq\Pi(0)=\frac{e\,g_\Zpr}{6\pi^2}\ln\frac{m_\tau}{m_\mu}\simeq 0.014\cdot g_\Zpr$, and the phenomenology of the $L_\mu-L_\tau$ model is similar to that of a dark photon model~\cite{Fabbrichesi:2020wbt}, with the substitution $\varepsilon = 0.0144\cdot g_\Zpr$, with $\varepsilon$ being the kinetic mixing parameter. At large $q^2$, $|\Pi(q^2)|$ grows, reaching a maximum value $\simeq 0.021 g_\Zpr$ at $q^2 = m^2_\mu$, and then smoothly goes to zero for $q^2\gg m^2_\mu$. 

\begin{figure*}[t]
    \centering
    \includegraphics[width=.9\textwidth]{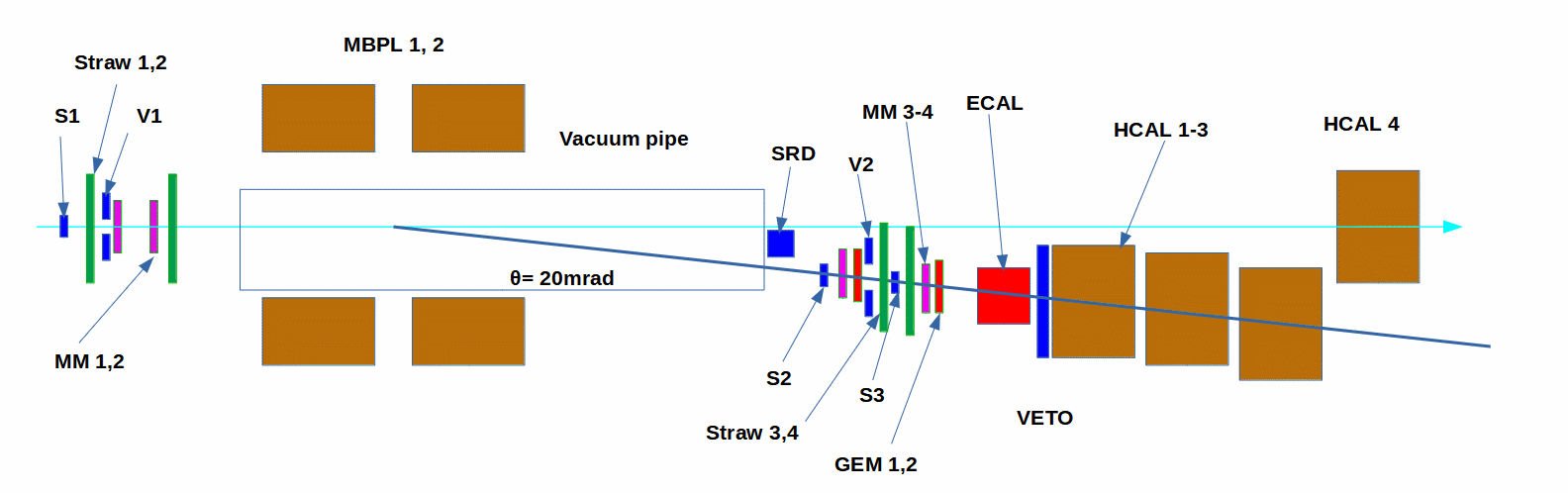}
    \caption{Schematic representation of the NA64$-e$ detector, searching  for an invisibly decaying $\Zpr$ produced by the interaction of 100 GeV electrons with the material of the active ECAL target. See text for further details.}
    \label{fig:detector}
\end{figure*}

\section{The $\Zpr$ search at NA64$-e$} Due to the loop-induced effective coupling, the $\Zpr$ paradigm can be explored at electron-beam experiments. In this work, we present the latest results of the search for the $\Zpr$ particle at the NA64$-e$ experiment at CERN SPS. 
In a collision of a high-energy electron beam with a thick target, $Z^\prime$ particles can be produced by two main processes involving the primary electron and the secondaries produced in the electromagnetic cascade, namely the radiative emission ($\Zpr$-strahlung) in the electromagnetic field of an atom by an electron/positron, $e^\pm N\rightarrow e^\pm N \Zpr$, and the resonant annihilation of a secondary positron with an atomic electron, $e^+e^-\rightarrow \Zpr$. Depending on the mass and on the couplings, the produced $\Zpr$ will then decay to one of the allowed dominant channels, i.e. $\mu^+\mu^-$, $\tau^+\tau^-$, $\nu\bar{\nu}$, or $\chi \chi$, where the neutrino channel refers both to $\nu_\mu$ and $\nu_\tau$. NA64$-e$ exploits the \textit{invisible} decay channel ($\nu\overline{\nu}$ or $\chi\chi$) to investigate the $\Zpr$ existence by relying on the missing-energy technique with a $E_0=$100 GeV electron beam impinging on an active thick target. The signature for the $Z^\prime$ production, followed by its invisible decay, is the observation of events
with a  well identified impinging $e^-$ track, in coincidence with a large missing energy $E_{miss}$, measured as the difference between the nominal beam energy $E_0$ and the one deposited in the active target $E_{ECAL}$, and no activity in the other downstream apparatus~\cite{Gninenko:2016kpg}. 
A schematic representation of the  NA64$-e$ detector assembly is shown in Fig.~\ref{fig:detector}. Incoming particles are detected by a set of three plastic scintillator counters (S1, S2, S3) and two veto counters (V1, V2). A magnetic spectrometer, consisting of tracking detectors (GEMs, MicroMegas, and Straw Tubes) placed upstream and downstream two dipole magnets (MBPL 1, MBPL 2) with a combined magnetic strength of approximately $7$~T$\cdot$m, is utilized for measuring particle momentum, with a resolution $\delta p/p$ of about $1\%$~\cite{Banerjee:2017mdu}. Particle identification is achieved by detecting synchrotron radiation (SR) photons emitted by the electrons deflected by the magnetic field through a compact Pb/Sc calorimeter (SRD)~\cite{Depero:2017mrr}. The NA64 active target is a Pb/Sc calorimeter (ECAL) with a thickness of 40 $X_0$, organized in a $5\times6$ matrix of $3.82\times3.82$ cm$^2$ cells. Each cell has independent PMT readout and is longitudinally segmented into a 4 $X_0$ pre-shower section (PRS) and a main section. Following the ECAL, a hermetic Fe/Sc hadron calorimeter (HCAL) with three modules, totaling approximately 21 $\lambda_I$ (nuclear interaction lengths), is installed to detect secondary hadrons and muons produced in the ECAL and upstream beamline elements. A fourth module is installed at zero degrees, to capture neutral particles produced by the beam interacting with upstream materials. A high-efficiency plastic scintillator counter (VETO) is placed between the ECAL and HCAL to further reduce background signals. 
The trigger for the experiment required the coincidence between signals from the scintillator counters, in anti-coincidence with the veto detectors, paired with an in-time energy deposition in the pre-shower $E_{PRS}\gtrsim 300$~MeV and in the whole ECAL $E_{ECAL} \lesssim 90$~GeV~\cite{Banerjee:2017hhz}.

\section{Data analysis} The results here presented are based on a total statistics of $ (9.1\pm0.5)\times10^{11}$ electrons on target (EOT) accumulated in different runs by NA64$-e$ during the 2016-2022 period, with beam intensity up to $\simeq 6\times 10^{6}$ electrons per SPS spill of 4.8 s. The data analysis was based on the same dataset already scrutinized and unblinded for the recently published invisibly-decaying dark photon search~\cite{Andreev:2023uwc}. Therefore, supported by the fact that the phenomenology of an invisible-decaying $\Zpr$ is very similar to that of the $\Apr$ search, we decided to adopt the same reconstruction algorithms and event selection criteria, resulting in zero observed events in the signal window~\cite{Banerjee:2017hhz}.  The criteria included the request for a well identified impinging electron track with momentum $100\pm10$~GeV, paired with an in-time energy deposition in the SRD detector in the range $\simeq 1-100$ MeV. Events with electro- and photo-nuclear interactions in the ECAL were suppressed by requiring the compatibility of the measured energy deposition in the different cells, in terms of longitudinal and transverse shape, with that expected from an electron-induced electromagnetic shower~\cite{Banerjee:2017hhz}. In addition, no activity must be observed in the VETO and in the HCAL detector; for the latter, a 1 GeV energy threshold was set, just above the noise level. Finally, we adopted the same value of the ECAL missing energy threshold defining the signal box ($E_{ECAL}<E^{thr}_{ECAL}$) that was optimized independently for each particular run period, accounting for small differences in the detector response and in the observed background levels. Obtained values were in the range $E^{thr}_{ECAL} \sim 47-50$~GeV.  The most critical background source originates from the interactions of the beam particles  with beamline materials upstream the ECAL. These interactions can produce secondary hadrons emitted at large angles, outside the detector acceptance, while the scattered  low-energy electron may hit the ECAL with reduced energy deposition, mimicking a signal~\cite{Andreas:2013lya,PhysRevD.94.095025}. To suppress this background, a multiplicity cut on the measured number of hits in the STRAW detectors was used. The residual yield was evaluated directly from the data, considering events lying in the sideband region $E_{HCAL}\simeq 0$, $E_{ECAL}<E_{0}$, and extrapolating the corresponding ECAL energy distribution in the signal region through an exponential model. For the 2016-2018 (2021-2022) runs, the expected background is $0.4\pm0.2 \,(0.3\pm0.2)$ events, where the error has been evaluated by varying the parameters of the model within their fit uncertainty and repeating the procedure. 
Secondary contributions arose from the in-flight decay of $\pi^-$ and $\mu^-$ beam contaminants~\cite{Andreev:2023xmj}, estimated with Monte Carlo, as well as from the punch-through of leading neutral hadrons $(n,K)$ produced in the ECAL, valuated from a dedicated measurement~\cite{Banerjee:2020fue}.


 \section{Signal simulation} \label{sec:sigsim}
 
\begin{figure}[t]
    \centering
    \includegraphics[width=.6\textwidth]{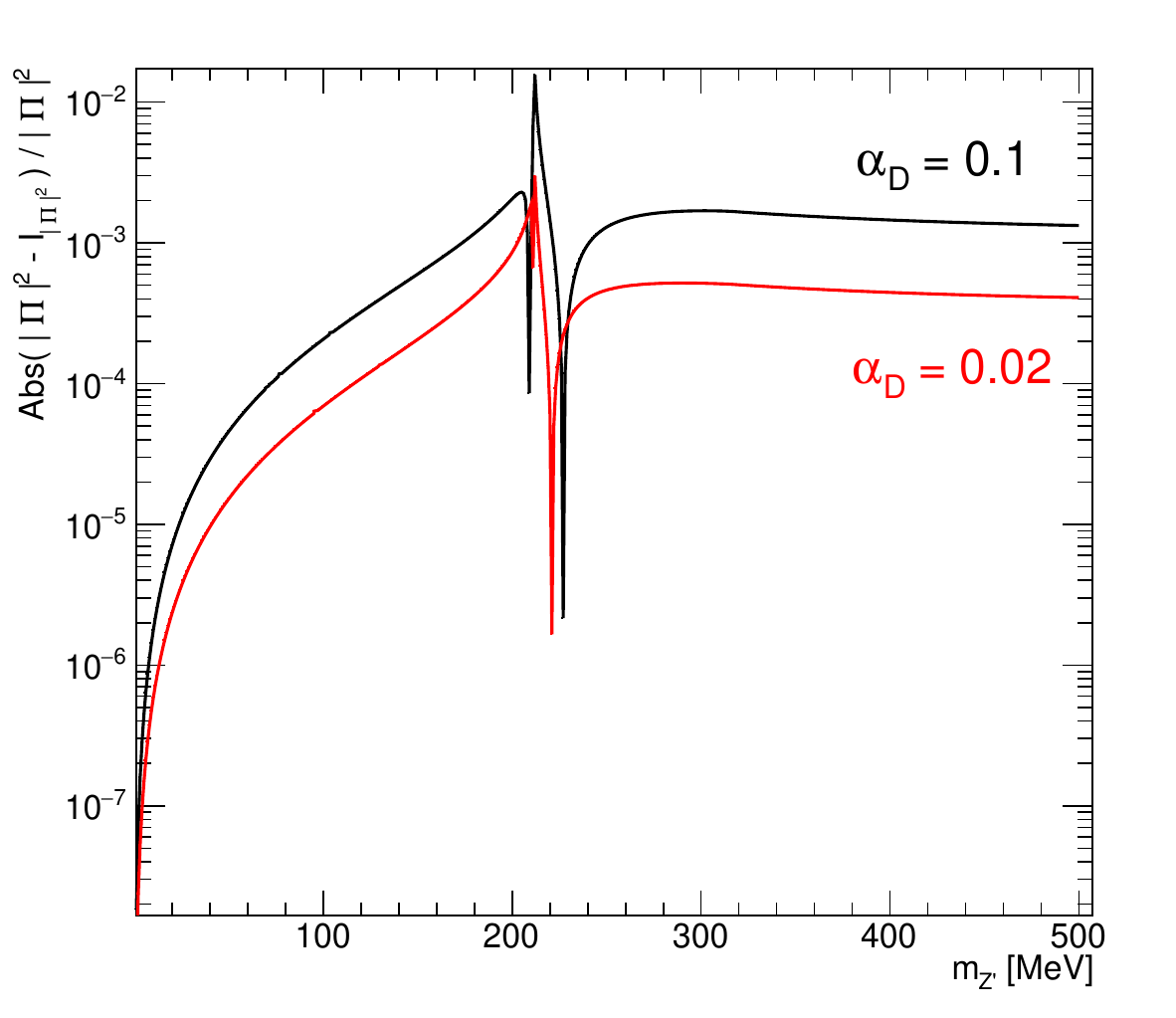}
    \caption{The relative difference between the $\Pi$ function on-shell value $|\Pi|^2$ and the $q^2$-averaged one $I_{|\Pi|^2}$ as a function of the $\Zpr$ mass, for different $\alpha_D$ values.}
    \label{fig:I_PI}
\end{figure}

 The expected yield $S$ of $\Zpr$ events within the signal window for given values of the model parameters was computed using the DMG4 software~\cite{Bondi:2021nfp,Oberhauser:2024ozf}, integrated within the full Geant4-based simulation of the NA64$-e$ detector~\cite{GEANT4:2002zbu,Allison:2016lfl}. We 
 considered two model variations: a ``vanilla'' case in which $g_D=0$ and a ``dark sector'' one featuring non-zero $Z^\prime-\chi$ DS coupling, focusing on the $1\,\mathrm{MeV}\lesssim m_\Zpr\lesssim500\,\mathrm{MeV}$ mass range for $g_\Zpr < 0.1$. For these parameters values, and in particular in the first scenario, the $\Zpr$ width is very narrow~\cite{NA64:2022rme}; this was appropriately treated in the simulation, as discussed in the following. Both the $\Zpr$-strahlung and the resonant annihilation production processes were implemented in DMG4, addressing the model peculiarities arising from the loop-induced $e^- - \Zpr$ coupling, as well the  $\Zpr$ width. We first considered the effects associated with the $q^2$ dependency of the $\Pi$ function, starting from the $\Zpr$-strahlung process. In the ``dark sector'' case and for the $\Zpr \rightarrow \chi \chi$ decay channel, the cross-section expression contains an average $\Zpr-e^-$ kinetic mixing value $I_{|\Pi|^2}$ (see also Ref.~\cite{NA64:2022rme}, Eq. A.5): 
\begin{equation}
I_{|\Pi|^2}=\int_{q^2_{min}}^{q^2_{max}}dq^2\frac{\left|\Pi(q^2)\right|^2}{\pi}\frac{\sqrt{q^2}\,\Gamma_\Zpr}{(q^2-m^2_\Zpr)^2+m^2_\Zpr\Gamma^2_\Zpr} \; ,
\end{equation}
where $q^2_{min}=4m^2_\chi$ and $q^2_{max}\gg m^2_\Zpr$ in the NA64$-e$ kinematic regime. In Fig.~\ref{fig:I_PI}, we compare $I_{|\Pi|^2}$ with the on-shell value $|\Pi(m^2_\Zpr)|^2$, for $\alpha_D=0.1$ (black curve) and $\alpha_D=0.02$ (red curve), by showing the relative difference among the two as a function of the $\Zpr$ mass\footnote{In the DS case, the width $\Gamma_\Zpr$ is dominated by the $\alpha_D$ contribution; we therefore neglected its dependency on $g_\Zpr$.}. As expected, the largest variation is seen for $m_\Zpr \simeq 2m_\mu$, since here the variation of the $\Pi$ function with respect to $q^2$ is larger. We observe that, as a consequence of the narrow $\Zpr$ width, the relative difference is always smaller than $\simeq 1\%$. This justifies the use of an on-shell approximation for the $\Zpr-$strahlung cross section, i.e. $I_{|\Pi|^2} \simeq |\Pi(m^2_\Zpr)|^2$. The same conclusion holds for the ``vanilla'' scenario, given the even narrower $\Zpr$ width. Finally, for the $e^+e^-$ resonant production the full $q^2$ dependency of the $\Pi$ function is naturally accounted for in the tree-level expression for $\sigma(s)$, where $s$ is the $e^+e^-$ invariant mass squared, given the identity $q^2=s$ -- see also Ref.~\cite{NA64:2022rme}, Eqs. (A.3) and (A.4). 

A second critical
effect considered in the simulation 
is the modification of the effective line shape for the $e^+e^-$ annihilation channel due to the atomic electrons motion. This effect is well known in atomic physics, where it induces, for example, a broadening of the photon energy spectrum in the $e^+e^-\rightarrow \gamma \gamma$ reaction~\cite{PhysRevB.61.10092}. In particular, for in-flight annihilation atomic motions manifest as the appearance of events with $E_\gamma > E_{max}$, where $E_{max}$ was the maximum allowed photon energy if the atomic electron was at rest~\cite{PhysRevLett.86.5612}. To our knowledge, this effect has never been implemented in a full Monte Carlo simulation of the resonant production of DS states, although recently an analytical calculation was performed in Ref.~\cite{Arias-Aragon:2024qji} (see Appendix~\ref{appendix}). Indeed, for a given positron energy $E_+$ in the reference frame of the target atom, the $e^+e^-$ invariant mass is given by the expression $s=2m_e^2+2E_+(E_--zP_-)$, being $E_-$ ($P_-$) the electron energy (momentum), and $z$ the cosine of the angle between the two leptons.
This implies that if the energy of a positron does not align with the resonance mass when interacting with a stationary electron, it can still do so when annihilating with an orbiting electron, and vice versa. 
\begin{figure}
    \centering
    \includegraphics[width=.62\textwidth]{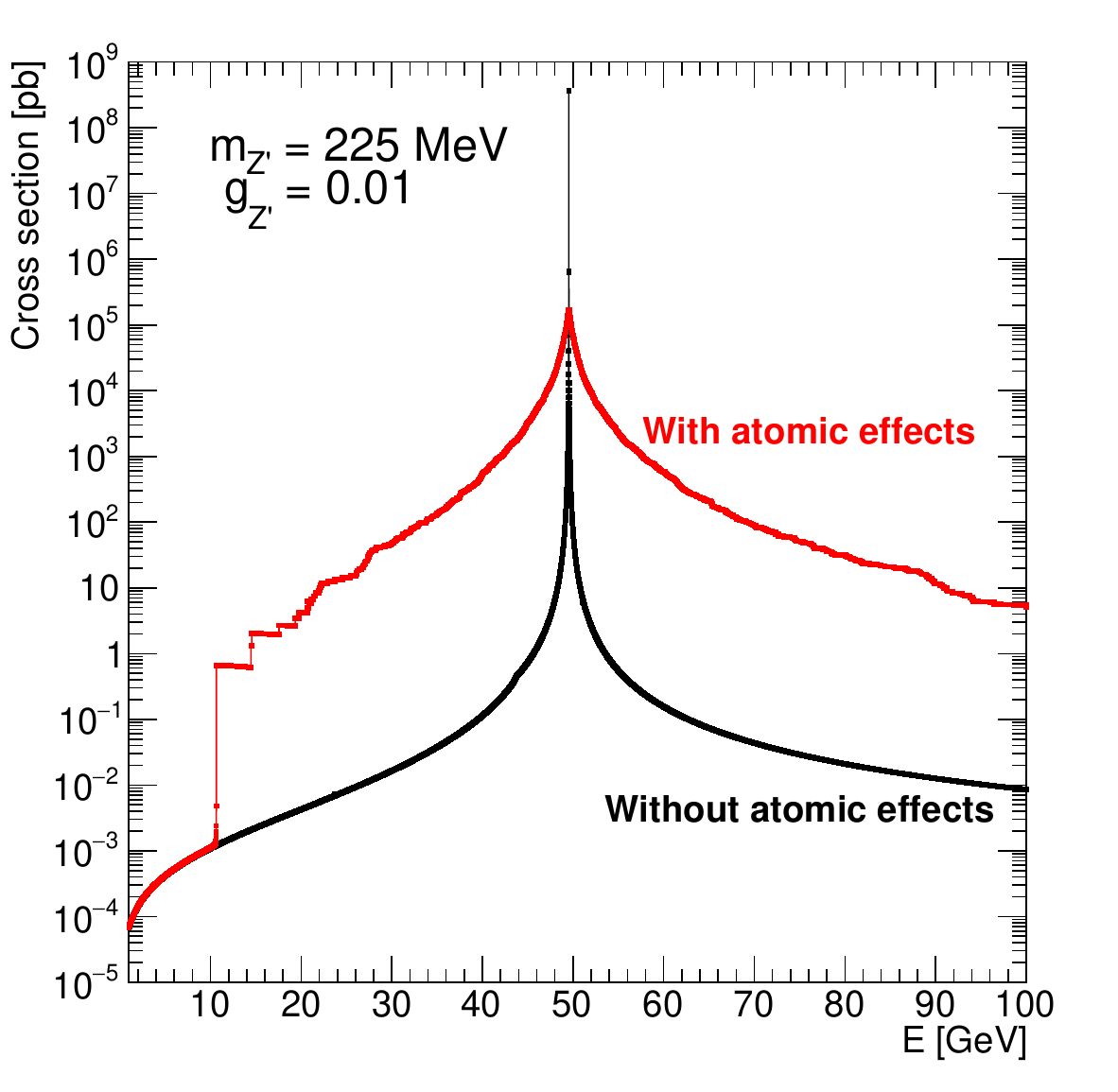}
    \caption{Total atomic cross section for the process $e^+e^-\rightarrow \Zpr \rightarrow \nu \bar{\nu}$ on lead as a function of the impinging positron energy, considering (red) or not (black) the motion of atomic electrons.}
    \label{fig:xs}
\end{figure}
To account for this, we computed an average cross section per atom, starting from the tree-level expression for the annihilation cross section, given by the formula
\begin{equation}
\sigma=\frac{A(E_+)}{(2m_e^2+2E_+(E_--zP_-)-m^2_
\Zpr)^2+m^2_\Zpr\Gamma_\Zpr^2}\;\;,
\end{equation}
where $A(E_+)$ is a smooth-varying function of the impinging positron energy. We assumed a flat distribution for $z$ and integrated over it to obtain an average value $\overline{\sigma}(E_+,E_-)$, depending solely on the positron and electron energy. Then, to integrate over $E_-$, we parameterized the kinetic energy distribution 
through a model inspired by the virial theorem, $\overline{T}=E_B$, where $-E_B$ is the binding energy, adopting an exponential ansatz $p(T_-)=\frac{1}{E_B}\exp(-T_-/E_B)$~\cite{PhysRevLett.86.5612}, and then we summed over the contributions from each atomic shell~\cite{Oberhauser:2024ozf}. For illustration, Fig.~\ref{fig:xs} shows the cross section per atom as a function of the positron energy for the reaction $e^+e^-\rightarrow \Zpr \rightarrow \nu \bar{\nu}$ on lead (``vanilla'' case) when $m_\Zpr=225$~MeV and $g_\Apr=0.03$, comparing the result obtained with (red) and without (black) the atomic effects. We observe that, due to atomic effects, the cross section value at the peak is reduced by many orders of magnitude, and simultaneously the shape is significantly enlarged, keeping a constant value for the integral with respect to the positron energy across the allowed kinematic range.

\begin{figure*}[t]
    \centering
    \includegraphics[width=.49\textwidth]{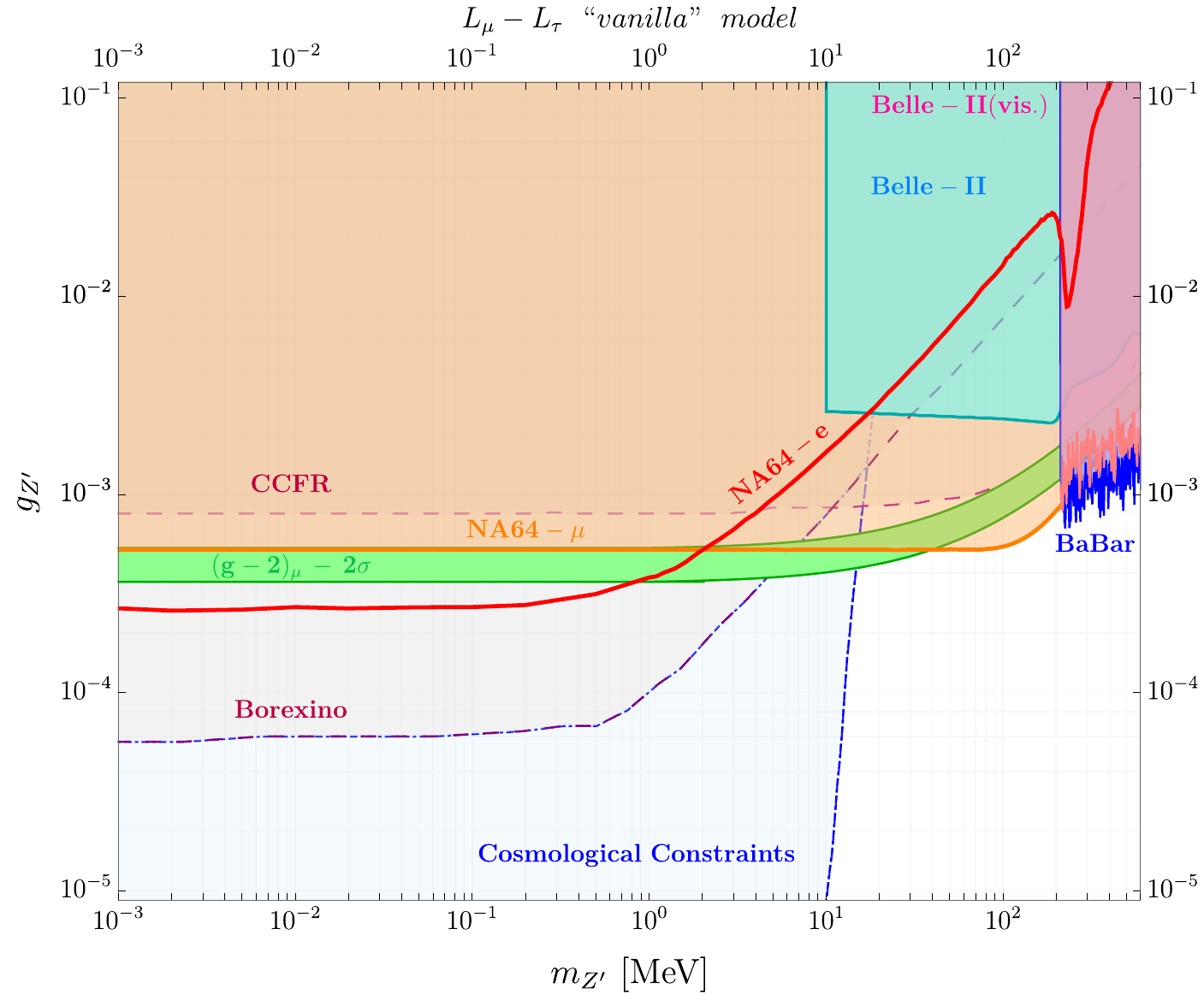}
    \includegraphics[width=.49\textwidth]{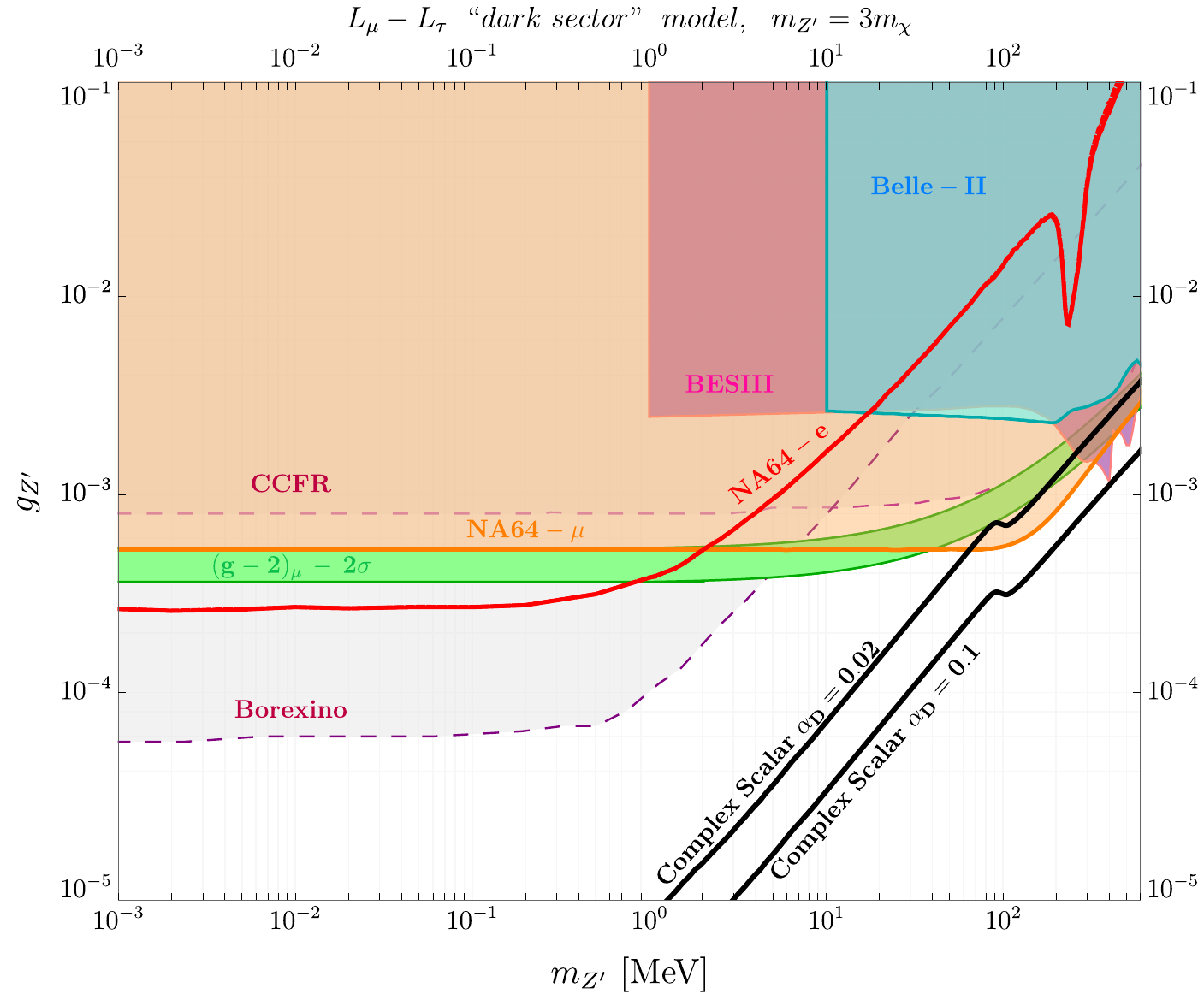}
    \caption{The new exclusion limits from the NA64$-e$ experiment in the $(g_\Zpr,m_\Zpr)$ parameter space for the $L_\mu-L_\tau$ model, considering the ``vanilla'' scenario without any DS coupling (left) and the full ``dark sector one'' (right), for $\alpha_D=0.1$ (continuous curve) and $\alpha_D=0.02$ (dashed curve).  The most competitive bounds reported by other experiments are also shown (see text for further details). The green band correspond to preferred combination of the model parameters that would solve the muon $g-2$ tension. The blue dashed line defines the region excluded by cosmological and astrophysical considerations. Finally, the black curves in the right plot are the preferred combination of the parameters to explain the observed dark matter relic density (so-called ``thermal target'').}
    \label{fig:Zp}
\end{figure*}

After the  $\Zpr$ yield evaluation via simulation, we estimated the signal efficiency $\kappa$ directly from the experimental data, applying the same selection cuts, except for the $E^{thr}_{ECAL}$ threshold, to pure impinging electron events measured during calibration runs. During these runs, no condition on the ECAL energy was included in the trigger, allowing to collect events where the 100-GeV electrons fully deposit their energy in the ECAL. 
The overall result reads $\kappa \simeq 50\%$, with a $\simeq 10\%$ variation between different runs mostly due to an increased beam intensity, and thus to a larger pile-up probability, for 2016 and 2022 periods. 
To account for the systematic uncertainty on the ECAL energy threshold, for each $Z^\prime$ mass value the procedure was repeated  multiple times by sampling $E^{thr}_{ECAL}$ from a Gaussian PDF, and then re-evaluating $S$. The RMS value of all results was taken as an estimate of $\Delta S$.
 Finally, an additional 15$\%$ uncertainty on the absolute yield was also introduced to account for the observed difference on the number of ``di-muon'' events between data and Monte Carlo~\cite{Andreev:2023uwc}.

\section{Results} Starting from the  zero events measured in the signal window, the predicted background level $B$, and the simulated $\Zpr$ yield, we computed an upper limit on the $g_\Zpr$ coupling as a function of $m_\Zpr$, for the cases $\alpha_D=0$ (``vanilla'') and $\alpha_D=0.02$ / $\alpha_D=0.1$ (``dark sector''). We used a frequentist method, considering the $90\%$ Confidence Level (CL) of a one-sided profile-likelihood test statistics~\cite{Gross:2007zz}. The likelihood model was built assuming a Poisson PDF for the number of events in the signal region, with mean value $\mu=S+B$. To handle the non-trivial dependency of $S$ on $g_\Zpr$, entering both in the cross-section multiplicative pre-factor and in the $\Zpr$ width~\cite{NA64:2022rme}, we proceeded by iteration, repeating at each time the computation of ${S}$ via Monte Carlo and the extraction of the limit by adopting the $g_\Zpr$ value returned from the previous iteration. The first iteration was performed using the $g_\Zpr$ values reported in our previous work~\cite{NA64:2022rme}. Convergence was observed already after three iterations.  

The obtained results are shown in Fig.~\ref{fig:Zp}, reporting in red the $90\%$~CL limit for the ``vanilla'' scenario (left plot) and for the ``dark sector'' one (right plot), with $\alpha_D=0.1$ (continuous line) and $\alpha_D=0.02$ (dashed line) and $m_\Zpr/m_\chi=3$. In the same plots, we also show recent exclusion limits from BaBar~\cite{BaBar:2016sci}, Belle-II~\cite{Belle-II:2022yaw} and BESIII~\cite{BESIII:2023jji}, as well as the latest result obtained by the NA64 collaboration through a dedicated missing-momentum experiment with a muon beam, NA64-$\mu$\cite{NA64:2024klw}. Results from previous neutrinos experiments (CCFR~\cite{Altmannshofer:2014pba} and Borexino~\cite{Kamada:2015era}) are also reported -- as already underlined in Ref.~\cite{Gninenko:2020xys}, these should be taken with care, being them a theoretical re-interpretation of the original experimental data. The blue dashed line defines the region excluded by cosmological and astrophysical considerations~\cite{Escudero:2019gzq}; this has been reported only in the ``vanilla'' case, since no connection with DM has been considered in~\cite{Escudero:2019gzq}. In both figures, the green band correspond to the preferred combination of the $\Zpr$ parameters that would solve the muon $g-2$ tension. Finally, in the ``dark sector'' scenario, the black lines report the predicted value of $g_\Zpr$ as a function of the $\Zpr$ mass from Eq.~\ref{eq:bound}. NA64-$e$ is the first accelerator-based experiment to unambiguously probe the $g-2$ band in the 1 keV $ < m_\Zpr \lesssim 2$ MeV mass range, confirming the re-analysis of the Borexino results. We notice that the limits set by NA64-$e$ tend to a constant value for a $\Zpr$ lighter than $\sim 0.1$ MeV. This is due to the interplay between the cross section of the $\Zpr$ radiative production and the  signal selection cuts in NA64-$e$. Due to the signal window definition, the experiment is sensitive only to  $\Zpr$s produced with a significant fraction of the initial $e^+/e^-$ energy $x \geq 0.5$. In this kinematic range,  if  $m_\Zpr \ll m_e$, the differential cross section for the $\Zpr$ radiative production is approximately independent from $m_\Zpr$~\cite{Gninenko:2017yus}, resulting in a flat limit
curve. 

\section{Conclusions}

In conclusion, we performed the analysis of the data collected by NA64-$e$ in the 2016-2022 period, searching for an invisibly decaying $\Zpr$ boson. In the data analysis, we adopted the same selection criteria identified in the search for an invisibly decaying Dark Photon, resulting in zero observed events in the signal box. The signal yield was evaluated by means of MC simulation: the dominant $\Zpr$ production mechanisms where implemented in the DMG4 software package, integrated in the Geant4-based MC framework of NA64-$e$, including the momentum-transfer dependency of the $\Zpr-e^-$ coupling and the effect of the motion of atomic electrons. 
Despite the disfavoured $\Zpr$ production processes with an electron beam, the obtained limits prove to be complementary to the NA64-$\mu$ constraints, excluding for the the first time with an accelerator-based experiment the $g - 2$ band in the low (1 keV - 2 MeV) $\Zpr$ mass region.

\appendix
\section{\label{appendix} Considerations on the effects of the motion of atomic electrons}

As described in the Sec.~\ref{sec:sigsim}, an effect to be considered in the simulation of the $e^+e^- \rightarrow \Zpr$ annihilation process is the modification of the effective annihilation line shape due to the motion of atomic electrons in the target. An analytical calculation of this effect,  assuming Compton profiles (CP) and Roothan-Hartree-Fock (RHF) wave functions for the modelization of the momentum distribution of the atomic shells, has been performed in~\cite{Arias-Aragon:2024qji}. The exclusion limits presented in this work have been obtained using the DMG4 simulation package, considering an effective exponential momentum distribution, based on the virial theorem, which allows a simple integration in a Monte Carlo simulation tool.
\begin{figure}[t]
    \centering
    \includegraphics[width=.85\textwidth]{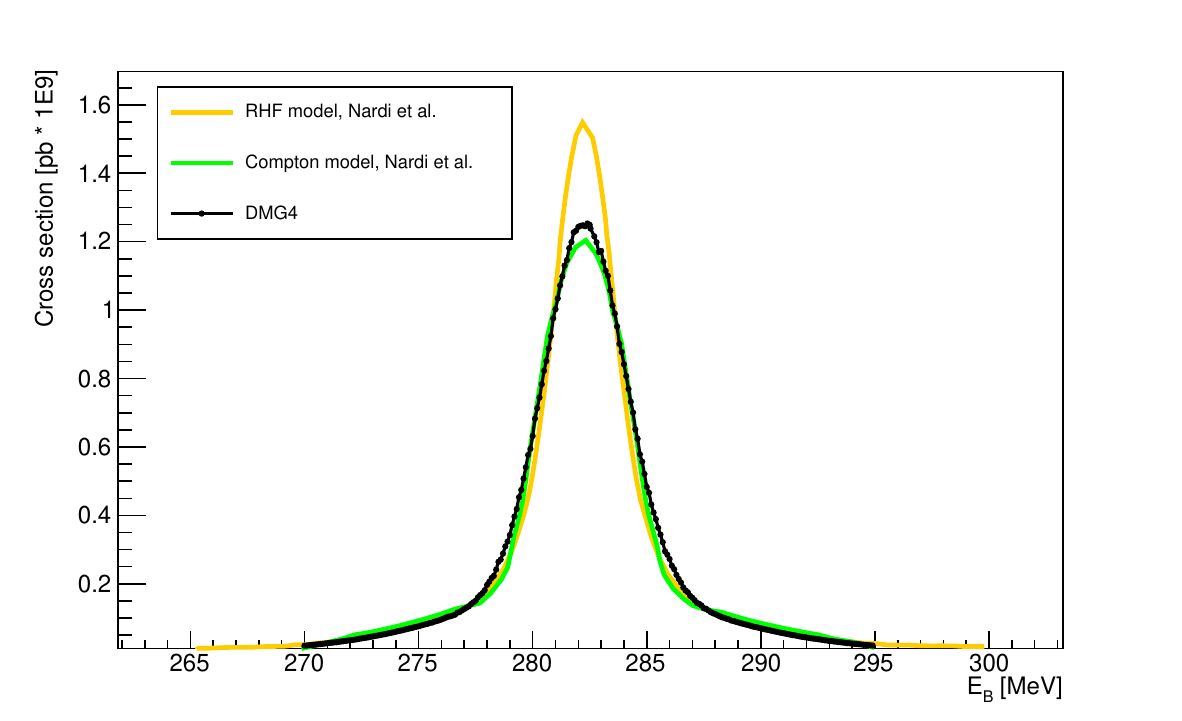}
    \caption{The three curves represent the cross sections of the resonant $\Zpr$ production on a thin diamond target as a function of the positron beam nominal energy $E_{B}$, obtained analytically considering the RHF (orange) and CP (green) model from~\cite{Arias-Aragon:2024qji}, and from a simulation via DMG4 (gray). See the text for more details.}
    \label{fig:comp}
\end{figure}
In order to check the agreement between the different approaches, we report in  Fig.~\ref{fig:comp} a comparison between the resonant cross section for a thin diamond target as a function of the beam energy  as reported in Fig.1 (left panel) of~\cite{Arias-Aragon:2024qji} and the same quantity evaluated with a DMG4-based simulation. The considered mass of the new vector boson is 17 MeV, while the beam energy spread is 0.5\%. We observe that DMG4 (gray curve) reproduces with very good accuracy the analytical calculation, with a maximum peak magnitude between the CP (green) and RHF (orange) curves.

\acknowledgments
We gratefully acknowledge the support of the CERN management and staff, and the technical staffs of the participating institutions for their vital contributions. 
This result is part of a project that has received funding from the European Research Council (ERC) under the European Union's Horizon 2020 research and innovation programme, Grant agreement No. 947715 (POKER). 
This work was supported by the HISKP, University of Bonn (Germany), ETH Zurich and SNSF Grant No. 186181, No. 186158, No. 197346, No. 216602 (Switzerland), and FONDECYT (Chile) under Grant No. 1240066, and ANID - Millenium Science Initiative Program - ICN2019 044 (Chile), and  RyC-030551-I and PID2021-123955NA-100 funded by MCIN/AEI/ 10.13039/501100011033/FEDER, UE (Spain), and COST Action COSMIC WISPers CA21106, supported by COST (European Cooperation in Science
and Technology. This work is partially supported by ICSC – Centro Nazionale di Ricerca in High Performance Computing, Big Data and Quantum Computing, funded by European Union – NextGenerationEU.

\bibliographystyle{apsrev4-2}
\bibliography{bibliographyNA64_inspiresFormat.bib,bibliographyNA64exp_inspiresFormat.bib,bibliographyOther_inspiresFormat.bib}

\end{document}